\documentclass[twocolumn,showpacs,prb,citeautoscript,superscriptaddress]{revtex4}
\usepackage[dvips]{graphicx}
\usepackage{dcolumn}
\usepackage{bm}
\usepackage{amsmath}
\usepackage{amssymb}
\usepackage{epsfig}
\usepackage{array}

\begin{document}
\title{Optical and thermodynamic properties of the high-temperature superconductor HgBa$_{2}$CuO$_{4+\delta}$}
\author{E. van Heumen}
\author{R. Lortz}
\author{A.B. Kuzmenko}
\author{F. Carbone}
\author{D. van der Marel}
\affiliation{University of Geneva, 24, Quai E.-Ansermet, Geneva 4,
Switzerland}
\author{X. Zhao}
\author{G. Yu}
\author{Y. Cho}
\author{N. Barisic}
\author{M. Greven}
\affiliation{Department of Physics, Applied Physics, and Stanford
Synchrotron Radiation Laboratory, Stanford University, Stanford,
CA 94305}
\author{C.C. Homes}
\affiliation{Condensed Matter Physics $\&$
 Materials Science Department, Brookhaven National
Laboratory, Upton, New York 11973, USA}
\author{S.V. Dordevic}
\affiliation{Condensed Matter Physics $\&$
 Materials Science Department, Brookhaven National
Laboratory, Upton, New York 11973, USA}
\affiliation{Current
adress: Department of Physics, The University of Akron, Akron, OH
44325, USA.}
\date{\today }

\begin{abstract}
In- and out-of-plane optical spectra and specific heat
measurements for the single layer cuprate superconductor
HgBa$_{2}$CuO$_{4+\delta}$ (Hg-1201) at optimal doping ($T_{c}$ =
97 K) are presented. Both the in-plane and out-of-plane superfluid
density agree well with a recently proposed scaling relation
$\rho_{s}\propto\sigma_{dc}T_{c}$. It is shown that there is a
superconductivity induced increase of the in-plane low frequency
spectral weight which follows the trend found in underdoped and
optimally doped Bi$_{2}$Sr$_{2}$CaCu$_{2}$O$_{8+\delta}$ (Bi-2212)
and optimally doped
Bi$_{2}$Sr$_{2}$Ca$_{2}$Cu$_{3}$O$_{10+\delta}$ (Bi-2223). We
observe an increase of optical spectral weight which corresponds
to a change in kinetic energy $\Delta W\approx$0.5 meV/Cu which is
more than enough to explain the condensation energy. The specific
heat anomaly is 10 times smaller than in
YBa$_{2}$Cu$_{3}$O$_{6+\delta}$ (YBCO) and 3 times smaller than in
Bi-2212. The shape of the anomaly is similar to the one observed
in YBCO showing that the superconducting transition is governed by
thermal fluctuations.
\end{abstract}

\maketitle

\section{Introduction}
In the 20 years after the discovery of high-temperature
superconductivity the determination of the generic properties of
these compounds has often been complicated by material and
crystallographic issues. Examples are the structural distortion in
Bi-2201, bi-layer splitting of the electronic bands due to
multiple copper-oxygen sheets per unit cell and the copper-oxygen
chains present in YBCO. In order to obtain information on the
phenomenon of superconductivity in the high $T_{c}$ cuprates that
is free from these complications one would like to study the
simplest possible structure. The mercury based cuprates with their
simple tetragonal structure and in particular Hg-1201, with only
one copper-oxygen sheet per unit cell and the highest $T_{c}$
($\approx$97 K) of the single layer compounds, seem to be good
candidates to achieve this goal. Moreover the critical
temperatures of these compounds are the highest obtained to date.
There are however some indications for intrinsic disorder in these
compounds\cite{sample}. These systems have not been as extensively
studied as other families because until recently sizeable single
crystals were lacking. In an earlier publication \cite{sample} the
successful growth of large single crystals of optimally doped,
single-layer Hg-1201 has been reported. These crystals have
subsequently been studied with resonant inelastic x-ray
scattering\cite{rixs} and ARPES\cite{ARPES}. In this paper we
present optical properties of one such a sample together with
measurements of the specific heat. The paper is organized as
follows: in section \ref{sample} some sample related issues are
discussed; section \ref{exp} explains the experimental techniques
used and in section \ref{res} the results are presented and
discussed. The conclusions are summarized in section \ref{conc}.

\section{Sample}\label{sample}
Large single crystals were grown using an encapsulation process as
discussed elsewhere\cite{sample}. The sample used for the in-plane
measurements is oriented with its largest surface along the a-b
plane with dimensions of $1.1 \times 1.4 \times 0.5$ mm$^{3}$.
Magnetic susceptibility measurements give a critical temperature
$T_{c}\approx$ 97 K with a somewhat broadened transition width of
5 K. The misorientation of crystallites is about
0.04$^{\circ}$\cite{sample}. Hg-1201 is highly hygroscopic and
does not cleave naturally along the ab-plane. Therefore we
polished the sample with a 0.1 $\mu$m diamond abrasive in a pure
nitrogen atmosphere before inserting it into the cryostat.

\section{Experiments}\label{exp}
\subsection{Optical measurements}\label{optics}
\begin{figure}[htb]
\includegraphics [scale = 0.7]{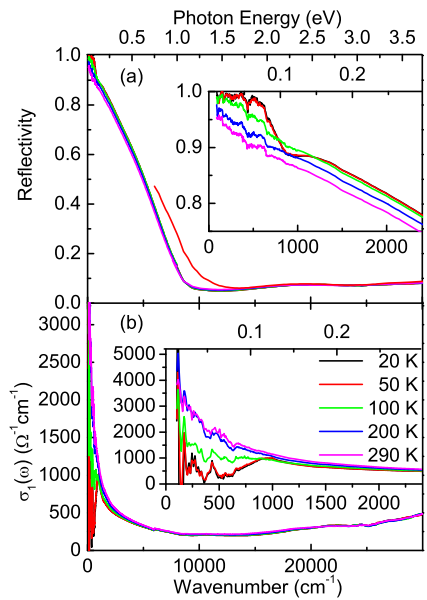}\\
\caption{\label{RandS}(Color online)(a): In-plane reflectivity for
selected temperatures. The reflectivity above 6000 cm$^{-1}$ has
been calculated from the measured dielectric function. The thin
red curve is the reflectivity at room temperature calculated from
the pseudo dielectric function. The inset shows the far infrared
reflectivity. (b): In-plane optical conductivity
$\sigma_{1}(\omega)$ for selected temperatures. The inset shows
the low frequency part on an expanded scale. Some sharp structures
in the region below 700 cm$^{-1}$ are due to the remnants of the
c-axis phonons. The temperatures are the same in all panels and
are indicated in the inset in fig \ref{RandS}b.}
\end{figure}

In-plane normal incidence reflectivity measurements have been
performed on a Fourier transform spectrometer in the frequency
range between 100 - 7000 cm$^{-1}$ (12 - 870 meV) coating the
sample \textit{in-situ} with gold to obtain a reference.
Ellipsometric measurements were made on a Woolam VASE ellipsometer
in the frequency range between 6000 and 30000 cm$^{-1}$ (0.75 -
3.72 eV) which yield directly the real and imaginary parts of the
dielectric function $\epsilon(\omega)$. The spectra are taken with
2 K intervals between 20 K and 300 K. Two home made cryostats are
used that stabilize the sample position during thermal cycling.
They operate at pressures in the order of $10^{-6}$ mbar in the
infrared region and $10^{-9}$ mbar in the visible light region.
Figure \ref{RandS}a shows the in-plane reflectivity
R$_{ab}$($\omega$) for selected temperatures. The reflectivity
curves above 6000 cm$^{-1}$ have been calculated from the
dielectric function. Also shown is the reflectivity calculated
from the pseudo dielectric function (see below).

\begin{figure}[htb]
\includegraphics [scale = 0.7]{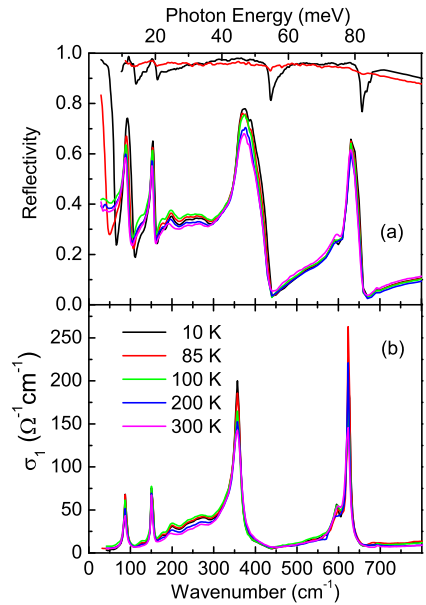}\\
\caption{\label{caxis}(Color online) (a): Far infrared c-axis
reflectivity at selected temperatures. Also shown is the in-plane
reflectivity at 20 K measured with (red) and without polarizer
(black). (b): The c-axis optical conductivity
$\sigma_{1}(\omega)$. The temperatures are the same in all panels
and are indicated in the inset in fig \ref{caxis}b.}
\end{figure}

In addition, the c-axis reflectivity R$_{c}$($\omega$) was
measured on the ac-plane of a different sample from 30 to 20000
cm$^{-1}$. Figure \ref{caxis}a shows the c-axis reflectivity for
selected temperatures. The c-axis optical conductivity is obtained
from a Kramers-Kronig (KK) transformation and is shown in figure
\ref{caxis}b.
In table \ref{DLparam} the oscillator parameters for the 5 phonon
lines are summarized. From a simple counting of atoms 8 phonon
modes are expected of which only four should be infrared active.
Comparing with other cuprates we attribute the first two phonon
lines to Hg and Ba vibrations, the mode at 355 cm$^{-1}$ to
vibration of the in-plane oxygens and the highest mode at 622
cm$^{-1}$ with the vibration of the apical oxygen mode. The fact
that this mode is split indicates that there is some oxygen
disorder present. The additional weak structure seen between 200
cm$^{-1}$ and 300 cm$^{-1}$ is probably also due to disorder.
Below $T_{c}$ a Josephson plasma edge appears that shifts with
temperature to a maximum of around 70 cm$^{-1}$ for the lowest
measured temperature of 10 K.

\begin{figure}[htb]
\includegraphics [scale = 0.7]{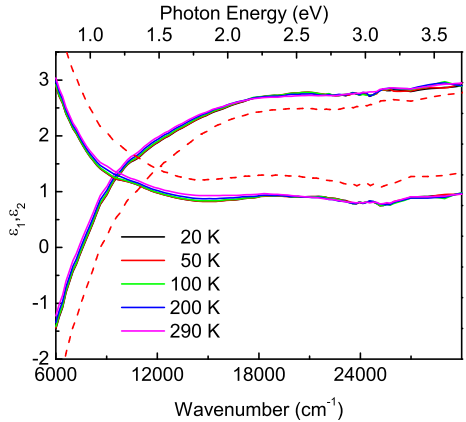}
\caption{\label{epsilon}(Color online) Dielectric function at
selected temperatures corrected for the c-axis contribution using
the method explained in the text. The pseudo dielectric function
at room temperature are indicated by the thin red curves.}
\end{figure}

The ellipsometric measurements were performed with an angle of
incidence of $60^{\circ}$. Due to the large angle of incidence the
measured pseudo dielectric function has an admixture of the c-axis
component. The c-axis dielectric function derived from the c-axis
reflectivity measurements was used to obtain the true in-plane
dielectric function $\epsilon_{ab}(\omega)$, by an inversion of
the Fresnel equations. The uncorrected dielectric function at room
temperature and corrected dielectric function for several
temperatures are shown in figure \ref{epsilon}.
Measurements were also performed for angles of incidence of
$65^{\circ}$, $70^{\circ}$ and $80^{\circ}$ and no dependence of
the corrected $\epsilon_{ab}(\omega)$ on the angle was found
showing the consistency of this procedure.

A disadvantage of using a cut and polished surface as compared to
a naturally cleaved one is the possible occurrence of a
misorientation of the crystal axes relative to the sample surface.
This results in the appearance of c-axis spectral features in the
in-plane reflectivity spectra\cite{Pimenov}. In figure
\ref{caxis}a, R$_{ab}$ measured with unpolarized light is shown. A
comparison with R$_{c}$ suggests that these features correspond to
c-axis phonons. Also shown is a spectrum where a polarizer was
oriented such that the spectral features observed in the first
spectrum are minimized. Using the model of
Ref[\onlinecite{Pimenov}] the misorientation is estimated to be of
the order of 3$^{\circ}$. In order to suppress the c-axis phonon
features in R$_{ab}$($\omega$), light polarized along the
intersection of sample surface and ab-plane is used for the
in-plane measurements. The features are not completely suppressed
however, likely due to the finite angle of incidence
($8^{\circ}$). This shows up in the optical conductivity and the
extended Drude analysis below.

\begin{figure}[htb]
\includegraphics [scale = 0.7]{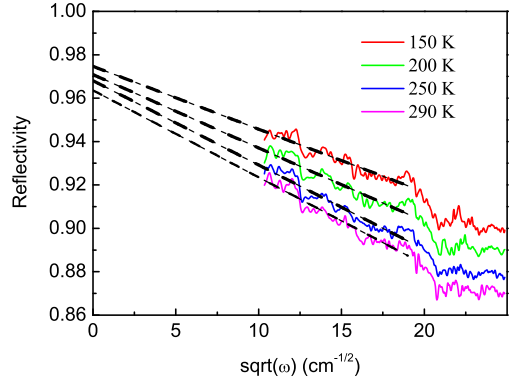}
\caption{\label{extrap}(Color online) Reflectivity plotted versus
$(\omega/2\pi c)^{1/2}$ and the fitted extrapolations (dashed) to
zero frequency.}
\end{figure}

Another experimental issue is related to the absolute value of the
in-plane reflectivity at low frequencies. In a metal for
frequencies $\omega\ll 1/\tau$ one can use the Hagen-Rubens
approximation\cite{wooten} to describe the reflectance:
\begin{equation}\label{HR}
R(\omega)\simeq
1-2\left(\frac{2\omega}{\pi\sigma_{0}}\right)^{1/2}.
\end{equation}
Here $\sigma_{0}$ is the dc conductivity. This is also expected to
approximately hold for cuprates in the normal state and this
approximation is frequently used to perform KK transformations of
reflectivity spectra. In figure \ref{extrap} the in-plane
reflectivity plotted versus $(\omega/2\pi c)^{1/2}$ is shown for
four different temperatures in the normal state.
The extrapolations have been obtained by fitting Eq. \ref{HR} to
the spectra between 100 cm$^{-1}$ and 300 cm$^{-1}$. One can see
that the reflectivity extrapolates to $\approx$ 0.97 instead of 1
for $\omega\to 0$. Since the curves all extrapolate to
approximately the same value the difference is presumably due to
the presence of a non-conducting secondary phase. From the fits we
extracted the dc conductivity and compared the temperature
dependence to resistivity measurements\cite{Barisic}. These two
measurements correspond in absolute value and follow the same
temperature dependence. However a problem arises when we try to
model R$_{ab}$ with a Drude-lorentz model. For temperatures below
$T_{c}$ the fits show no sign of superconductivity. Instead of a
delta function or very narrow Drude like contribution we find a
rather broad Drude peak. Moreover the temperature dependence of
the dc conductivity obtained from the model does not correspond
with the resistivity measurements. We find that if we scale the
in-plane reflectivity measurements upwards by 3$\%$ for all
temperatures the dc conductivity extracted from the model closely
follows the dc resistivity measurements and the Drude peak below
$T_{c}$ is replaced by a delta function.

To extract the optical conductivity from the measured in-plane
reflectivity and dielectric function a KK constrained variational
analysis is used\cite{kuz1}. First, R$_{ab}$($\omega$) and
$\epsilon_{ab}(\omega)$ are fitted simultaneously to a standard
Drude-Lorentz model of which the parameters are given in table
\ref{DLparam} for the room temperature spectra. Note that the last
oscillator falls outside the measured spectral range.
\begin{table}[htb]
\begin{tabular}{rrrrrrrr|rrrrrrr}
\hline\hline
$\epsilon_{\infty}$ && $\omega_{o}$ && $\omega_{p}$ && $\Gamma$ && $\omega_{o}$ && $\omega_{p}$ && $\Gamma$\\
\hline 2.53 && 0 && 1.5 & & 0.13 && 86 && 150 && 10\\
&& 0.48 && 1.34 && 0.9 & & 151 && 140 && 5\\
&& 1.41 && 0.44 && 0.67 & & 355 && 475 && 22\\
&& 2.51 && 1.54 && 1.78 & & 595 && 300 && 35\\
&& 3.36 && 0.23 && 0.24 & & 622 && 305 && 10\\
&& 4.87 && 3.92 && 2.21 & & && &&\\
\hline\hline
\end{tabular}
\caption{Left: Oscillator parameters of the in-plane room
temperature Drude-Lorentz model in eV (except $\epsilon_{\infty}$
which is dimensionless). Right: c-axis phonon parameters in
cm$^{-1}$}\label{DLparam}
\end{table}
In a second step this model is then refined using multi-oscillator
variational dielectric functions that fit all spectral details of
R$_{ab}$($\omega$) and $\epsilon_{ab}(\omega)$. This approach
improves the determination of the high frequency optical
conductivity because the measured dielectric function in the
visible light region is used to anchor the unknown phase of the
reflectivity. The in-plane reflectivity measurements have been
scaled upwards by 3$\%$ as explained above.

In figure \ref{RandS}b the optical conductivity is shown for
selected temperatures. It displays many of the features common to
cuprates. At room temperature the low frequency spectrum is
dominated by a Drude like peak that narrows when temperature is
decreased. Below $T_{c}$ a partial gap opens up followed by an
onset in absorption that starts around 500 - 600 cm$^{-1}$. This
value for the onset is comparable to optimally doped Bi-2212, YBCO
and Tl-2201\cite{Timusk} which all have approximately the same
critical temperature but very different crystal structures
indicating that this onset originates in the CuO$_{2}$ planes.
Above 0.5 eV the optical conductivity is only weakly temperature
dependent. We find evidence for at least three interband
transitions with resonance energies of 1.41 eV, 2.51 eV and 3.36
eV. The latter two transitions have resonance energies close to
resonances observed in the RIXS study of Ref[\onlinecite{rixs}].
Unfortunately, the c-axis contamination of the in-plane spectra
prevents us from making rigorous statements about any in-plane
phonon features.

\subsection{Specific heat measurements}\label{SHsec}
Specific-heat measurements have been performed in magnetic fields
up to 14 T on a small piece broken from the crystal used for the
optical measurements with a mass of $\sim$700 $\mu$g. The magnetic
fields used in these measurements are oriented perpendicular to
the Cu-O planes. Since the specific heat is a probe of bulk
thermodynamic properties the measurements confirm the bulk nature
of the superconducting transition. It also gives an estimate of
the strength of thermal fluctuations and of the superconducting
condensation energy. A micro-relaxation calorimeter adapted to the
small size of the crystal was used. Data has been taken using a
generalized long relaxation method as described
elsewhere\cite{Wang}. This method gives a high precision in the
determination of absolute values and a sensitivity comparable to
that of standard AC methods as used in previous publications on
Hg-1201\cite{Carr1,Billon}. The absolute error is estimated as ~5
\% due to the mass of the thermal compound used to mount the small
sample. This does not enter into relative measurements when data
taken in a magnetic field is used as a background.

\begin{figure}[htb]
\includegraphics [scale = 0.7]{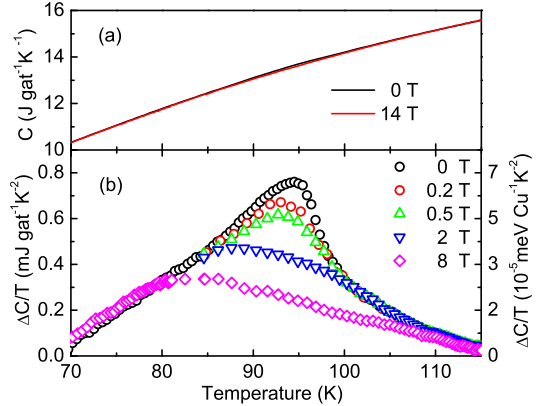}
\caption{\label{Specheat}(Color online) Difference of the specific
heat data $\Delta C/T=(C(B,$T$)-C(14 T,$T$))/T$ for a number of
magnetic fields $B$ showing the anomaly at the superconducting
transition.}
\end{figure}

In figure \ref{Specheat}a we show two representative specific heat
curves measured in fields of 0 T and 14 T.
Figure \ref{Specheat}b shows the difference $\Delta C/T$ between
measurements taken in a field $B$ and that taken in a field of 14
T in $C/T$. Due to the much higher upper critical field,
subtracting data taken in a 14 T field does not provide the exact
compensation of the phonon background but helps to investigate the
shape and size of the anomaly. Note that the anomaly is only 0.4
$\%$ of the total specific heat and thus particularly small as
compared to other cuprate superconductors.

\section{Results}\label{res}
\subsection{Extended Drude analysis}\label{extdrudesec}
The original Drude theory of non-interacting electrons assumes a
frequency independent scattering rate $1/\tau=const$. This
assumption does not hold in systems where the charge carriers
interact with a bosonic spectrum or where strong correlations are
important. This theory was extended by Allen and Mikkelsen to
include a frequency dependent scattering rate $1/\tau(\omega)$ in
which the optical conductivity takes the form\cite{Allen},
\begin{equation}
\sigma(\omega,T)=\frac{1}{4\pi}\frac{\omega^{2}_{p}}{1/\tau(\omega,T)-i\omega
m^{*}(\omega,T)/m_{b}},
\end{equation}
with $\omega_{p}$ the bare plasma frequency of the free charge
carriers, $m^{*}$ the effective mass and $m_{b}$ the band mass.
Note that since $\sigma(\omega,T)$ is causal, $1/\tau(\omega,T)$
and $m^{*}(\omega,T)/m_{b}$ obey KK relations. Although
$1/\tau(\omega,T)$ and $m^{*}(\omega,T)/m_{b}$ are simply related
to the real and imaginary parts of 1/$\sigma(\omega)$, there is a
subtlety that one has to consider to reliably extract these two
quantities from the data. This is best illustrated by expressing
$1/\tau(\omega,T)$ and $m^{*}(\omega,T)/m_{b}$ in terms of the
dielectric function
$\epsilon(\omega)=\epsilon_{1}(\omega)+i\epsilon_{2}(\omega)$:
\begin{equation}
\frac{1}{\tau(\omega)}=\frac{\omega^{2}_{p}}{\omega}\frac{\epsilon_{2}(\omega)}{[\epsilon_{\infty,IR}-\epsilon_{1}(\omega)]^{2}+\epsilon_{2}^{2}(\omega)}
\end{equation}
and
\begin{equation}
\frac{m^{\*}(\omega)}{m_{b}}=\frac{\omega^{2}_{p}}{\omega^{2}}\frac{\epsilon_{\infty,IR}-\epsilon_{1}(\omega)}{[\epsilon_{\infty,IR}-\epsilon_{1}(\omega)]^{2}+\epsilon_{2}^{2}(\omega)}.
\end{equation}
Here $\epsilon_{\infty,IR}$ is the contribution to the dielectric
function in the infrared arising from interband transitions which
should not be taken into account in the single component approach.
The subtlety is in the choice of $\epsilon_{\infty,IR}$ and
$\omega_{p}$. The choice of $\epsilon_{\infty,IR}$ is not so
important at low energies where
$|\epsilon_{1}|\gg\epsilon_{\infty,IR}$ but becomes important at
higher energies. For example, early studies of the cuprates
indicated that the scattering rate was linear in frequency up to
energies as high as 1 eV\cite {Azrak}, but if
$\epsilon_{\infty,IR}$ is chosen as below the scattering rate
starts to show signs of saturation already around 0.5
eV\cite{Dirk}. Here the following convention is adopted:
$\epsilon_{\infty,IR}=\epsilon_{\infty}+\sum_{j}S_{j}$ where the
$S_{j}=\omega_{p,j}/\omega_{0,j}$ are the oscillator strengths of
the interband transitions obtained from a Drude-Lorentz fit (see
Table \ref{DLparam}) and $\omega_{p}$ is chosen such that
$m^{*}(\omega,T)/m_{b}$ approaches unity at high energy ($\sim$1
eV). This leads to the choice $\epsilon_{\infty,IR}\approx3.6$ and
$\omega_{p}\approx$16500 cm$^{-1}$ in the present case. The value
of $\epsilon_{\infty,IR}$ is slightly smaller than the ones found
in Ref[\onlinecite{Dirk}] for Bi-2212, likely due to the smaller
volume fraction of oxygens in the unit cell. The contribution to
the dielectric function from the polarizability of oxygen is
easily calculated using the Clausius-Mossotti relation,
\begin{equation}\label{Claussius}
\epsilon_{\infty,IR}\approx 1+\frac{4\pi
N\alpha/V}{1-\frac{4\pi}{3}N\alpha/V}=
1+\frac{\alpha_{0}}{1-\gamma\alpha_{0}}
\end{equation}
where $N$ is the number of oxygens per unit cell, V is the unit
cell volume and $\alpha$ is the polarizability of the oxygen
atoms. In the second equality we have defined
$\alpha_{0}\equiv4\pi N\alpha/V$. The factor $\gamma=1/3$
appearing in the denominator is a geometrical factor for a cubic
crystal structure which we keep for simplicity. For O$^{2-}$ the
ionic polarizability $\alpha\approx$3.88$\cdot$10$^{-24}$
cm$^{3}$. Using unit cell parameters for Hg-1201 of $a\times
b\times c$ = 3.85 $\times$ 3.85 $\times$ 9.5 $\dot{A}$ and 4
oxygen atoms per unit cell we find
$\epsilon_{\infty,IR}\approx3.56$, which is fortuitously close to
our estimate of 3.6 in view of the lesser agreement for Bi-2212
and Bi-2223 as we now show. For optimally doped Bi-2212 and
optimally doped Bi-2223 we use $a\times b\times c$ = 5.4 $\times$
5.4 $\times$ 30.8 $\dot{A}$ and $a\times b\times c$ = 5.4 $\times$
5.4 $\times$ 37.1 $\dot{A}$ respectively. These are the cell
parameters corresponding to 4 formula units, i.e. $N=32$ and
$N=40$ oxygens respectively. This yields
$\epsilon_{\infty,IR}\approx5.16$ and
$\epsilon_{\infty,IR}\approx5.52$. These values are substantially
larger than the estimated experimental values
$\epsilon_{\infty,IR}\approx4.5$ for Bi-2212\cite{Dirk} and
$\epsilon_{\infty,IR}\approx4.1$ for Bi-2223.
\begin{figure}[htb]
\includegraphics [scale = 0.7]{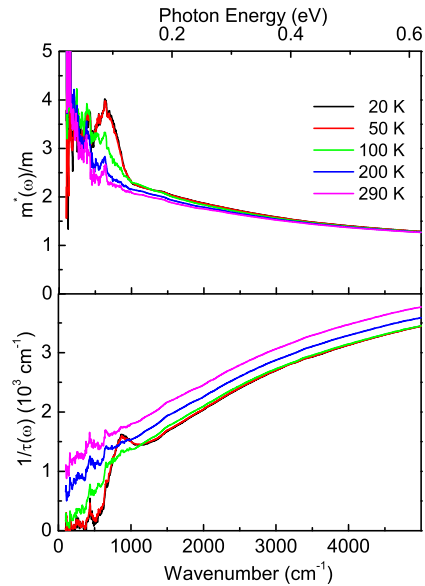}
\caption{\label{taumstar}(Color online) The frequency dependent
scattering rate and effective mass of Hg-1201. The small peak in
$m^{*}(\omega,T)/m_{b}$ at $\approx$ 620 cm$^{-1}$ for
temperatures above $T_{c}$ is due to the remnant phonon
structure.}
\end{figure}
Figure \ref{taumstar} shows $1/\tau(\omega)$ and
$m^{*}(\omega)/m_{b}$ for selected temperatures. The scattering
rate is strongly suppressed below 600 cm$^{-1}$ for temperatures
below $T_{c}$ indicative of the opening of a gap. ARPES
measurements on Hg-1201 show a maximum gap value of 30 meV, but
there is some uncertainty in this value because no quasiparticle
peak is observed around the anti-nodal direction\cite{ARPES}. From
the optical measurements it is difficult to extract the gap value:
simple s-wave BCS superconductors in the dirty limit show an onset
in absorption associated with the superconducting gap at
2$\Delta$. This onset is shifted to $2\Delta+\Omega_{ph}$, with
$\Omega_{ph}$ a phonon resonance energy, if the coupling to
phonons is included. It has been suggested that the onset seen in
cuprates is shifted due to the interaction of the electrons with
the magnetic resonance mode\cite{Hwang} by $\Omega_{mr}$ to
$2\Delta+\Omega_{mr}$. The onset here is defined as the point
where the rise in $1/\tau(\omega)$ is steepest. This corresponds
by KK relations to a maximum in $m^{*}(\omega)/m_{b}$. Using the
values $\Omega_{mr}\approx 52 \pm 20$ meV\cite{sample} and
$\Delta_{0}\approx 30 \pm 10$ meV\cite{ARPES} we find
$2\Delta+\Omega_{mr}\approx 110 \pm 30$ meV ($900\pm 240$
cm$^{-1}$), which is of the same order of magnitude as our
experimental value of $80\pm5$ meV.
\begin{figure}[htb]
\includegraphics [scale = 0.7]{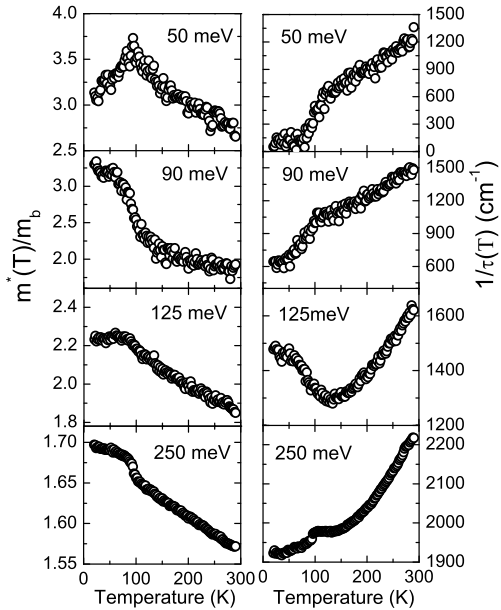}
\caption{\label{extdrudetdep}Temperature dependence of
$1/\tau(\omega,T)$ and $m^{*}(\omega,T)/m_{b}$ for selected
energies. Both quantities show a clear departure from the normal
state trend at $T_{c}$.}
\end{figure}
In figure \ref{extdrudetdep} the temperature dependence of
$1/\tau(\omega,T)$ and $m^{*}(\omega,T)/m_{b}$ is shown for
selected photon energies. In the normal state, the temperature
dependence of $1/\tau(\omega,T)$ for small photon energies is
linear in temperature but this linearity is lost for photon
energies larger than $\hbar\omega>0.1$ eV. The scattering rate
shows a sharp drop at $T_{c}$ for all photon energies except in a
narrow window between 95 and 140 meV where the scattering rate
increases below $T_{c}$. It has a maximum increase around 110 meV.
The temperature dependence of $m^{*}(\omega,T)/m_{b}$ becomes
roughly linear above $\hbar\omega>0.1$ eV and shows an increase
when the system becomes superconducting over most of the frequency
range, except below 55 meV where it decreases.

\subsection{Superfluid density}\label{scdensity}
\begin{figure}[htb]
\includegraphics [scale = 0.7]{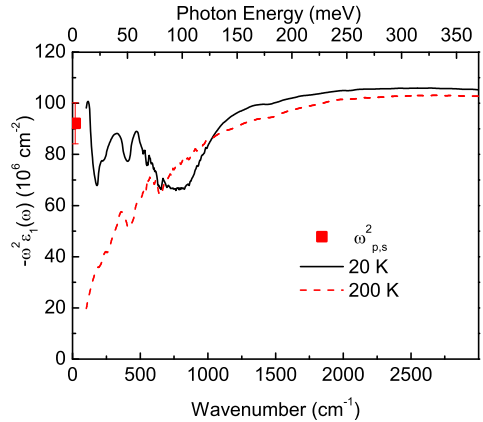}
\caption{\label{wsqeps}$-\omega^{2}\epsilon_{1}(\omega)$ plotted
at 20 K and 200 K . Extrapolating the 20 K curve to $\omega=0$
agrees with the value $\omega_{p,s}$ = 9600 $\pm$ 400 cm$^{-1}$
found from the Drude-Lorentz fit of the spectrum indicated by the
square at $\omega$= 0.}
\end{figure}

The in-plane condensate strength $\rho_{s}=\omega_{p,s}^{2}$ is
determined in two ways. The first method is to fit the low
temperature spectrum with a Drude-Lorentz model. In such a model
the superfluid density is represented by a $\delta$-peak at zero
frequency with strength $\omega_{p,s}^{2}$. The value obtained
from this method is $\omega_{p,s}$ = 9600 $\pm$ 400 cm$^{-1}$
($1.2\pm 0.05$ eV). The error bar on this quantity is determined
by shifting the reflectivity up and down by 1 $\%$ and observing
the subsequent change in $\omega_{p,s}$. The corresponding values
for Bi-2212 and Bi-2223 are $\omega_{p,s}$ = 9500 cm$^{-1}$ and
$\omega_{p,s}$ = 10300 cm$^{-1}$ respectively\cite{fabri}. The
second method relies on the assumption that the real part of the
dielectric function in the superconducting state is dominated at
low frequencies by the superfluid density,
\begin{equation}\label{epsapprox}
\epsilon_{1}(\omega)\approx-\frac{\omega_{p,s}^{2}}{\omega^{2}}
\end{equation}
Since $\epsilon_{1}(\omega)$ has been determined in a model
independent way this result is more reliable. In figure
\ref{wsqeps} the function $-\omega^{2}\epsilon_{1}(\omega)$ is
shown. The extrapolation to zero frequency of this function
matches well with the value obtained with the Drude-Lorentz fit.

It is interesting to use the value for $\omega_{p,s}$ extracted
from the in-plane measurements and from c-axis measurements to
check the scaling relation
$\rho_{s}\propto\sigma_{dc}T_{c}$\cite{homes}. This has already
been done for the c-axis measurements in Ref. [\onlinecite{homes}.
From a Drude-Lorentz fit to the c-axis data a value of
$\omega_{p,s}\approx290\pm10$ cm$^{-2}$ is found, more than a
factor of 30 different from the in-plane value. At the same time
from an extrapolation $\omega\to 0$ we find
$\sigma_{1,c}(\omega\to 0,T\simeq T_{c})\approx$ 7
$\Omega^{-1}$cm$^{-1}$ for the c-axis and $\sigma_{1,ab}(\omega\to
0,T\simeq T_{c})\approx$8300 $\Omega^{-1}$cm$^{-1}$. Put together
we find $N_{c}\simeq 4.6\sigma_{dc}T_{c}$ and $N_{ab}\simeq
4.3\sigma_{dc}T_{c}$ in good agreement with the scaling trend
observed in Ref[\onlinecite{homes}].

\subsection{Low frequency spectral weight}\label{SW}
In recent years there has been a lot of interest in the
temperature dependence of the low frequency integrated spectral
weight,
\begin{equation}\label{SWeq}
W(\Omega_{c},T)=\int_{0}^{\Omega_{c}}\sigma_{1}(\omega,T)d\omega\approx\frac{\pi
e^{2}a^{2}}{2\hbar^{2}V}<-\widehat{T}>,
\end{equation}
where \textit{a} is the in-plane lattice constant, \textit{V} the
unit cell volume and $\widehat{T}$ the kinetic energy operator.
$\Omega_{c}$ is a cut off frequency chosen to approximately
separate the intraband  and interband transitions. In a nearest
neighbor tight binding model the relation between
W($\Omega_{c}$,T) and $<-\widehat{T}>$ is exact. More recently it
has been shown that this relation still holds approximately in the
doping range under consideration\cite{frank}. According to BCS
theory the kinetic energy increases when the system is driven into
the superconducting state but it theoretically
\cite{hirsch,anderson,eckl,wrobel,haule,toschi,marsiglio,mayer,norman}
and experimentally\cite{basov,hajo,syro,kuz2,fabri} it was found
that in cuprates the kinetic energy decreases over a large doping
range. Recently, in studies of the doping dependence of Bi-2212 it
was found that on the overdoped side of the phase diagram the
kinetic energy follows the BCS prediction\cite{deutscher,fabri2}.
Experimentally, $W(\Omega_{c},T)$ gives a handle on the
superconductivity induced change in kinetic energy. A qualitative
indication of the superconductivity induced changes of low
frequency spectral weight can be obtained from the dielectric
function measured directly with ellipsometry. This can be done by
monitoring the shift of the screened plasma frequency
$\omega_{p}^{*}=\omega_{p}/\sqrt{\epsilon_{\infty}}$, \textit{i.e}
the frequency for which $\epsilon_{1}\left(\omega\right)=0$, with
temperature. Although it gives a first indication it does not give
a definite answer since $\omega_{p}^{*}$ can be influenced by
other factors, for instance the temperature dependence of the
interband transitions.
\begin{figure}[htb]
\includegraphics [scale = 0.7]{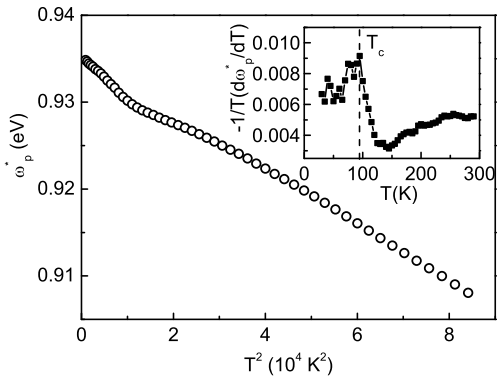}
\caption{\label{plasmaf}Screened plasma frequency $\omega_{p}^{*}$
versus T$^{2}$. The relative error bars on this quantity are
comparable to the size of the points. The inset shows the
derivative.}
\end{figure}
In figure \ref{plasmaf}, $\omega_{p}^{*}$ is plotted versus
$T^{2}$. The screened plasma frequency shows an extra blueshift
below $T_{c}$, suggesting an increase of the low frequency
spectral weight. Note that the absolute value of $\omega_{p}^{*}
(\approx$ 7300 cm$^{-1}$ at room temperature) is smaller than the
one observed for optimally doped Bi-2212\cite{hajo} ($\approx$
7600 cm$^{-1}$) and Bi-2223\cite{fabri} ($\approx$ 8100 cm$^{-1}$)
which can be related to a smaller volume density of CuO$_{2}$
planes. In contrast to earlier observations $\omega_{p}^{*}$ shows
a deviation from the $T^{2}$ behavior already in the normal state.

\begin{figure*}[htb]
\includegraphics [scale = 0.7]{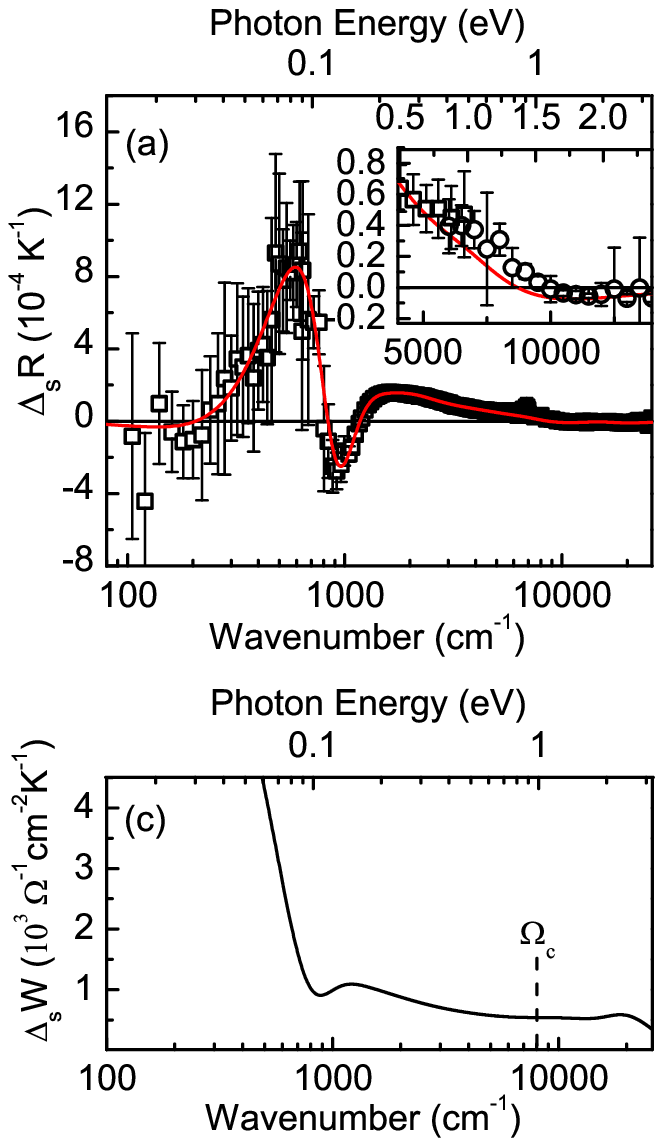}%
\includegraphics [scale = 0.7]{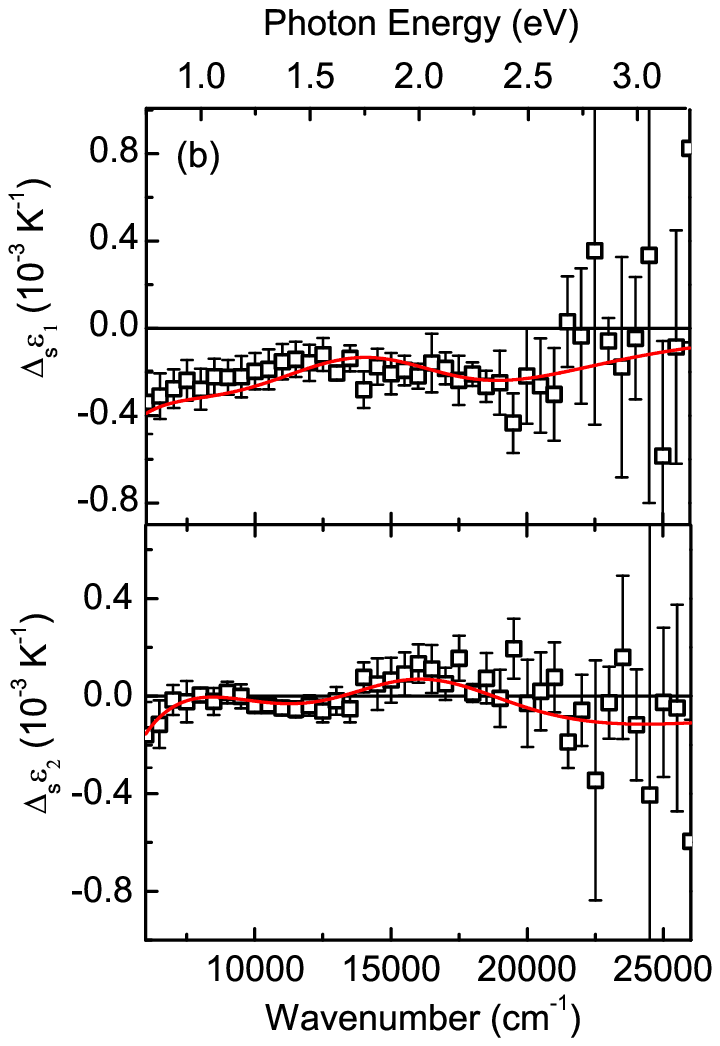}
\caption{\label{slopeanalysis}Spectral dependence of (a):
$\Delta_{s}R$ (squares), (b): $\Delta_{s}\epsilon_{1,2}$
(squares). Solid lines in (a) and (b) are the fits of the
$\Delta_{s}R$ and $\Delta_{s}\epsilon_{1,2}$ spectra. The inset in
(a) shows the excellent match between the directly measured
$\Delta_{s}R$ (squares) and the $\Delta_{s}R$ calculated from
$\Delta_{s}\epsilon_{1,2}$ (circles). (c): $\Delta_{s}W$
calculated from the best fit to $\Delta_{s}R$ and
$\Delta_{s}\epsilon_{1,2}$ using Eq. \ref{deltaW}. The observed
spectral weight increase is $\Delta_{s}W(\Omega_{c})= +540$
$\Omega^{-1}$cm$^{-2}K^{-1}$. The dashed line indicates the cut
off frequency used to calculate W($\Omega_{c}$,T).}
\end{figure*}

To calculate the integral in Eq. \ref{SWeq} one formally has to
integrate from $\omega=0$. Therefore to do the integration one has
to rely on an extrapolation of the reflectivity to zero frequency.
It has recently been shown that the superconductivity induced
increase of spectral weight and absolute value of
$W(\Omega_{c},T)$ are rather insensitive to this extrapolation,
provided one also uses the information contained in
$\sigma_{2}(\omega,T)$\cite{fabri,kuz3}.
\begin{figure}[htb]
\includegraphics [scale = 0.7]{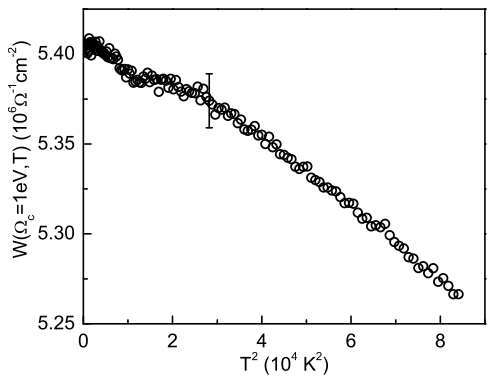}
\caption{\label{sw1ev}(Color online) Integrated spectral weight
for a cut off $\Omega_{c}$ = 1 eV. The error bar indicates the
estimated error due to the extrapolations used.}
\end{figure}
Figure \ref{sw1ev} shows the integrated spectral weight using Eq.
\ref{SWeq} with the cut off $\Omega_{c}$ = 1 eV. One would like to
choose $\Omega_{c}$ such that it separates the free carrier
response from the bound charge response. This is strictly speaking
not possible because in most cuprates the two regions overlap.
Therefore the cut off is chosen in a region where the
superconductivity induced change of spectral weight is more or
less constant (figure \ref{slopeanalysis}c). By extrapolating the
normal state trend to $T=0$, the superconductivity induced
increase of spectral weight is estimated to be $\Delta
W\approx1.5\pm1\cdot10^{4}$ $\Omega^{-1}$ cm$^{-2}$ which is 0.5
$\%$ of the total spectral weight. A remark has to be made here on
the temperature dependence of $W(\Omega_{c},T)$. The temperature
dependent reflectivity measurements show a hysteresis between
curves measured cooling down and curves measured warming up for
frequencies between 3000 cm$^{-1}$ and 7000 cm$^{-1}$. This
hysteresis is probably caused by a small amount of gas absorption
on the sample surface. This has the effect of suppressing the
reflectivity in this range below 150 K with a maximum suppression
at 20 K of about 0.5 $\%$. This trend was not observed in the
ellipsometric data in the region of overlap. The upward kink at
$T_{c}$ is not affected by this, but the deviation from $T^{2}$
behavior below 150 K could be a result of this. Note that
$W(\Omega_{c},T)$ closely follows $\omega_{p}^{*}(T)$, so the
temperature dependence of $W(\Omega_{c},T)$ is probably not too
much affected. In several other cuprates the temperature
dependence of the normal state optical spectral weight is
quadratic. Here we observe this quadratic temperature dependence
only for temperatures higher than 170 K. The coefficient of this
quadratic part has been observed to be unexpectedly large and is
believed to be due to correlation effects\cite{toschi}. To compare
the superconductivity induced increase of $W(\Omega_{c},T)$ to
other compounds we express $\Delta W$ in meV/Cu. We find $\Delta
W\approx$0.5 meV/Cu. This is a factor of 2 smaller than observed
for Bi-2212\cite{hajo} and a factor of 4 smaller than the one for
Bi-2223\cite{fabri}.

\begin{figure}[htb]
\includegraphics [scale = 0.7]{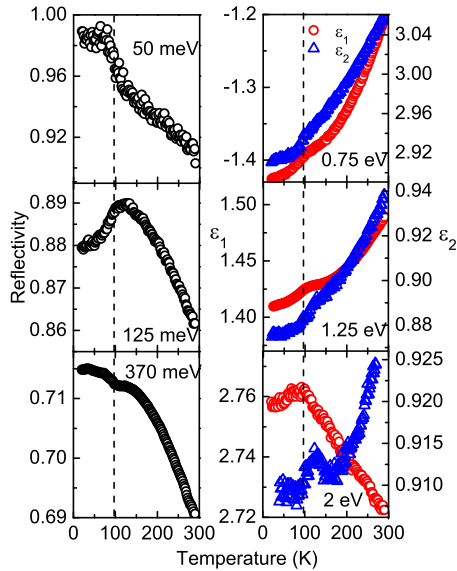}
\caption{\label{tdep}Temperature dependencies of the in-plane
reflectivity and dielectric function for selected energies. The
dashed lines indicate $T_{c}$.}
\end{figure}

In order to separate the superconductivity induced changes from
the temperature dependence of the normal state it is necessary to
measure the changes in the optical constants as a function of
temperature. Because the superconducting transition is second
order, the superconductivity induced changes appear as a kink at
$T_{c}$ as can be seen in figure \ref{tdep}.
A determination of the superconductivity induced spectral weight
transfer by analysis of slope changes at $T_{c}$ is more reliable
since this method is much less sensitive to systematic errors in
the absolute value of reflectivity. The details of this analysis
are explained in Ref.[\onlinecite{kuz2}]. The slope change at
$T_{c}$ for any optical quantity \textit{f($\omega$,T)} is defined
as\cite{kuz2},
\begin{equation}\label{fderiv}
\Delta_{s}f(\omega)=\left(\frac{\partial f(\omega,T)}{\partial
T}\right)_{T_{c}+\delta}-\left(\frac{\partial
f(\omega,T)}{\partial T}\right)_{T_{c}-\delta}.
\end{equation}
Since the slope changes are smeared due to fluctuations, the
derivatives correspond to average values in certain regions
$\delta$ above and below $T_{c}$. For this analysis one needs
temperature dependencies with a dense sampling in temperatures.
Figure \ref{tdep} shows the measured temperature dependencies for
selected photon energies. From these temperature dependencies the
slope change at $T_{c}$ can be estimated. Since numerical
derivatives of the data are very noisy, the temperature
dependencies are fitted above and below $T_{c}$ to a second order
polynomial\cite{kuz2}. The derivatives are then simply calculated
from the analytical expressions. Figure \ref{slopeanalysis} shows
the slope change of the reflectivity and the dielectric function
at $T_{c}$.
The functions $\Delta_{s}R(\omega)$,
$\Delta_{s}\epsilon_{1}(\omega)$ and
$\Delta_{s}\epsilon_{2}(\omega)$ were obtained by fitting the data
(figure \ref{tdep}) between 40 K and $T_{c}$ (= 97 K) and between
$T_{c}$ and 200 K. The error bars were estimated by varying the
upper and lower limits for both the high and low temperature fits.

From the inset of figure \ref{slopeanalysis}a one can see the good
agreement between the slope change estimated from reflectivity and
ellipsometric measurements. Spectra of $\Delta_{s}\epsilon_{1}$
and $\Delta_{s}\epsilon_{2}$ satisfy KK relations, therefore
$\Delta_{s}\epsilon$ can be modelled with an oscillator model that
has the same functional form as a Drude-Lorentz model,
\begin{equation}
\Delta_{s}\epsilon=\Delta_{s}\epsilon_{\infty}+\sum_{i=0}^{N}\frac{A_{i}}{\omega_{i}^{2}-\omega^{2}-i\gamma_{i}\omega}
\end{equation}
where $\Delta_{s}\epsilon_{\infty}$ is the change in the high
energy contribution, $\omega_{i}$ is the center frequency of the
\textit{i}th oscillator, $\gamma_{i}$ is its width and A$_{i}$ is
the oscillator strength. The difference with a normal
Drude-Lorentz model is that the oscillator strength can take on
both positive and negative values corresponding to the addition or
removal of spectral weight. $\Delta_{s}\sigma(\omega)$ is easily
calculated from $\Delta_{s}\epsilon$ and the slope difference
integrated spectral weight is then defined as,
\begin{equation}\label{deltaW}
\Delta_{s}W(\Omega_{c})=\frac{A_{0}}{8}+\int_{0+}^{\Omega_{c}}\Delta_{s}\sigma_{1}(\omega)d\omega
\end{equation}
with $A_{0}$ the condensate strength. This quantity is displayed
in figure \ref{slopeanalysis}c. The result gives
$\Delta_{s}W(\Omega_{c})= +540$ $\Omega^{-1}$cm$^{-2}K^{-1}$ for a
cut off $\Omega_{c}\approx$ 1 eV. This result is somewhat lower
than that for Bi-2212 ($\Delta_{s}W(\Omega_{c})= +770$
$\Omega^{-1}$cm$^{-2}K^{-1}$)\cite{kuz2} and Bi-2223
($\Delta_{s}W(\Omega_{c})= +1100$
$\Omega^{-1}$cm$^{-2}K^{-1}$)\cite{fabri} but has the same sign
and order of magnitude. It is also possible to calculate
$\Delta_{s}W(\Omega_{c})$ directly from the measured temperature
dependence of $W(\Omega_{c},T)$, by fitting $W(\Omega_{c},T)$ in
exactly the same way as is done for the other optical quantities.
The result obtained in this way is
$\Delta_{s,direct}W(\Omega_{c})= +420\pm$150
$\Omega^{-1}$cm$^{-2}K^{-1}$ which is a little bit lower but
consistent with the result obtained using the temperature
modulation analysis.

\subsection{Specific heat anomaly}
Figure \ref{Specheat}b shows specific-heat data taken in different
magnetic fields after subtracting the 14 Tesla data as a
background. The specific heat in zero field clearly shows an
asymmetric lambda shape typical for a superconducting transition
governed by thermal fluctuations of the anisotropic 3D-XY
universality class\cite{Meingast,Schneider}. The shape of the
anomaly shows more resemblance to that of optimally doped
YBCO\cite{Junod1} than to that of the more two dimensional Bi-2212
which shows a symmetric anomaly\cite{Junod2}. As in
YBCO\cite{Junod1}, the peak is clearly shifted to lower
temperatures in a magnetic field. This shift is well described in
the framework of 3D-XY fluctuations\cite{Schneider}. A
contribution due to fluctuations can be seen up to $\sim$15 K
above $T_{c}$. Since a 14 T field is by far not enough to suppress
all pair correlations above $T_{c}$, there probably also is a
contribution at higher temperatures which can not be observed with
this method\cite{Meingast,Lortz}. The influence of a magnetic
field on the fluctuations is to introduce a magnetic length
$l=(\Phi_{0}/B)^{1/2}$ closely related to the vortex-vortex
distance which reduces the effective dimensionality
\cite{Lortz,Schneider}. In the critical 3D-XY regime, the
correlation length diverges upon approaching $T_{c}$ following a
power law of the form $\xi\sim\xi_{0}(1-t)^{-\nu}$ with
$t=T/T_{c}$ and $\nu$=0.67 for the 3D-XY universality class. The
presence of a magnetic length scale cuts off this divergence
within a certain temperature window and thus broadens the
transition in a field as seen in the specific heat data. The
superconducting transition is thus far from being a BCS type
transition where the transition temperature is reduced in a field
because the Cooper pairs are broken up. Here $T_{c}$ is rather a
phase-ordering transition comparable to that of superfluid 4He,
albeit with a clear anisotropy. Due to the lack of the exact
phonon background it is not possible to estimate up to which
temperatures fluctuations due to phase correlations of preformed
Cooper pairs exist. However, it should be noted here that no
anomaly at higher temperatures is visible, suggesting a rather
continuous evolution of pair correlations at higher temperatures
with a continuous opening of a superconducting gap far above
$T_{c}$, whereas the lifetime of phase fluctuations is
finite\cite{Orenstein}. An interesting point is that a field of 14
T has a clear influence on the fluctuations above $T_{c}$ (between
100 K and 115 K), while no effect of an applied field has been
reported for YBCO\cite{Junod1}. This suggests a stronger pair
breaking effect by a magnetic field and thus points to a smaller
upper critical field in this compound. In a BCS type
superconductor the specific heat in the normal and the
superconducting state is used to evaluate the thermodynamic
critical field $H_{c}(T)$ which is a measure of the condensation
energy. In the presence of fluctuation contributions it has been
argued that fluctuation contributions above $T_{c}$ have to be
considered to calculate the real condensation energy\cite{Dirk2}.
As the true phonon background is unknown here this calculation
cannot be performed. Information can be obtained by comparing the
size of the specific heat anomaly with that of other cuprate
superconductors having a similar $T_{c}$. Nevertheless the
specific heat jump  $\Delta C=0.052$ J gat$^{-1}$ K$^{-1}$ can be
compared to $\Delta C=0.2$ J gat$^{-1}$ K$^{-1}$ found for
optimally doped Bi-2212\cite{Junod2} and $\Delta C=0.46$ J
gat$^{-1}$ K$^{-1}$ for overdoped YBCO\cite{Junod1} (the
abbreviation gat stands for gramatom). The large difference to the
value found in YBCO can be explained by the larger contribution of
fluctuation above $T_{c}$ in Hg-1201 (and Bi-2212) which appears
to have a larger anisotropy and thus more 2D correlations above
$T_{c}$. A large part of the condensation energy can thus be
attributed to the smooth onset of 2D correlations of Cooper pairs
far above $T_{c}$. The anomaly which is visible at $T_{c}$
represents thus only a small part of the condensation
energy\cite{Dirk2}. If the units eV/K and data normalized per
Cu-atom are used, the single layer compound can be compared to the
2-layer compound Bi-2212. This leads to $\Delta
C$=4.3$\cdot10^{-3}$ meV K$^{-1}$ per copper for Hg-1201 which is
3.5 times smaller than the value found for optimally doped Bi-2212
$\Delta C$=1.5$\cdot10^{-2}$ meV K$^{-1}$ per copper.

\begin{figure}[htb]
\includegraphics [scale = 0.7]{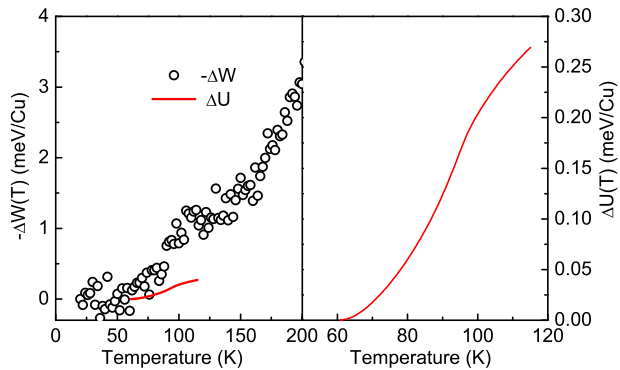}
\caption{\label{intE}Comparison between the kinetic energy
relative to 20 K, -$\delta
W(T)=W(\Omega_{c},T)-W(\Omega_{c},T=20K)$ on the left and the
internal energy $\Delta U(T)$ relative to 60 K on the right. The
red line in the left panel is $\Delta U(T)$ plotted on the same
scale as $\delta W(T)$.}
\end{figure}

From the specific heat we estimate the change in internal energy
by integrating,
\begin{equation}
\Delta U(T)=\int_{0}^{T}[C(0T,T')-C(14T,T')]dT'
\end{equation}
Figure \ref{intE} shows a comparison between the internal energy
relative to 60 K and the kinetic energy relative to 20 K defined
as -$\delta W(T)=W(\Omega_{c},T)-W(\Omega_{c},T=20K)$.
We find that $\delta W(T)$ is about 3 times larger than the
$\Delta U(T)$, so in principle $\delta W(T)$ is large enough to
account for the condensation energy. We note again that the change
in internal energy is probably only a small part of the
condensation energy because the fields used here are by far not
enough to completely suppress superconductivity. The temperature
dependence of $\delta W$ around $T_{c}$ is much more gradual than
the one for $\Delta$U. Because of the better signal-to-noise ratio
of the specific heat data it is possible to observe the change in
internal energy due to the phase ordering transition as a rather
sharp step at $T_{c}$. Superimposed on this is the much more
gradual change due to the 2d fluctuations. Since the phase
ordering transition occurs in a narrow window around $T_{c}$ and
contributes only a small part of the change in internal energy, we
are probably not able to observe this change in $\delta W(T)$.

\section{Conclusions}\label{conc}
Optimally doped, single layer  HgBa$_{2}$CuO$_{4+\delta}$ has been
investigated using optical and calorimetric techniques. From an
extended Drude analysis we find a reasonable agreement between the
onset in absorption as seen in the optical measurements and the
reported values for the gap and the resonance energy of the
magnetic mode. We have presented detailed temperature dependencies
of $m^{*}(\omega,T)/m_{b}$ and $1/\tau(\omega,T)$. The in-plane
superfluid plasma frequency is found to be
$\omega_{p,s}\approx9600\pm400$ cm$^{-1}$ whereas the out-of-plane
$\omega_{p,s}\approx290\pm10$ cm$^{-1}$. Both the in-plane and
out-of-plane superfluid densities fall on the scaling relation of
Ref[\onlinecite{homes}]. From an analysis of the temperature
dependent spectral weight W($\Omega_{c}$,T) we find that the low
frequency spectral weight shows a superconductivity induced
\textit{increase}. The corresponding decrease in kinetic energy
$\Delta W\approx$0.5 meV per copper is sufficient to explain the
condensation energy extracted from the specific heat measurements.
We observe a specific heat anomaly that is comparable in shape to
YBCO and shows a similar field dependence, but has a size of the
anomaly is 10 times smaller however.

\section{Acknowledgments}
We gratefully acknowledge stimulating discussions with F.
Marsiglio. The work at the University of Geneva is supported by
the Swiss National Science Foundation through the National Center
of Competence in Research "Materials with Novel Electronic
Properties-MaNEP". Work at Brookhaven was supported by the Office
of Science, U.S. Department of Energy, under Contract No.
DE-AC02-98CH10886. The crystal growth work at Stanford University
was supported by the DOE under Contract No. DE-AC02-76SF00515 and
by the NSF under Grant No. 0405655.

\end{document}